\documentstyle[eqsecnum,aps]{revtex}

\begin{document}
\draft \preprint{HEP/123-qed}
\title{Mass Renormalization in the Su-Schrieffer-Heeger Model}
\author{Marco Zoli}
\address{Istituto Nazionale Fisica della Materia - Universit\'a di Camerino, \\
62032 Camerino, Italy. e-mail: zoli@campus.unicam.it }

\date{\today}
\maketitle
\begin{abstract}
This study of the one dimensional Su-Schrieffer-Heeger model in a
weak coupling perturbative regime points out the effective mass
behavior as a function of the adiabatic parameter
$\omega_{\pi}/J$, $\omega_{\pi}$ is the zone boundary phonon
energy and $J$ is the electron band hopping integral. Computation
of low order diagrams shows that two phonons scattering processes
become appreciable in the intermediate regime in which zone
boundary phonons energetically compete with band electrons.
Consistently, in the intermediate (and also moderately
antiadiabatic) range the relevant mass renormalization signals the
onset of a polaronic crossover whereas the electrons are
essentially undressed in the fully adiabatic and antiadiabatic
systems. The effective mass is roughly twice as much the bare band
value in the intermediate regime while an abrupt increase (mainly
related to the peculiar 1D dispersion relations) is obtained at
$\omega_{\pi}\sim \sqrt{2}J$.
\end{abstract}
\pacs{PACS: 63.10.+a, 63.20.Dj, 71.38.+i}
\narrowtext
%\widetext
A sizeable electron-phonon interaction can induce a local
deformation in the lattice accompanied by the formation of a
quasiparticle with multiphononic character, the polaron
\cite{landau}. As the spatial extension of the lattice deformation
can vary, the concepts of large and small polaron have been
introduced \cite{toyozawa,emin,devreese,mott,mello,fehske}: the
transition between a large and a small polaron state is driven by
the strength of the electron-phonon coupling
\cite{rashba,raedt,kopida,tsiro} and monitored through the
behavior of ground state properties such as the polaron energy
band and the effective mass \cite{romero,kornilo,jeckel}. In
particular, an abrupt increase of the polaron mass versus the {\it
e-ph} coupling  is  associated with the occurence of the
self-trapping event and the loss of the polaron mobility
properties. The aforementioned investigations of polarons
generally depart from the Hamiltonian of the Holstein model
 \cite{holst59} in which the electron couples
to dispersive optical phonons \cite{io1} through a local (momentum
independent) interaction while the coupling to acoustical phonons,
although possible in principle \cite{io2}, would lead to huge mass
renormalizations \cite{farias}. In fact, also the Holstein optical
polaron masses are very heavy (at least larger than $10^3$ times
the bare band mass) in the self-trapped state but masses of order
$10$ times the bare band mass are possible in the presence of high
energy phonon spectra \cite{io3}.

As an alternative to the Holstein molecular crystal model with
local interactions one may consider the Su-Schrieffer-Heeger (SSH)
model Hamiltonian \cite{ssh}. The tight binding SSH model  was
introduced to study polyacetylene \cite{ssh1}, an essentially one
dimensional  polymer with  delocalized $\pi$ electrons which are
responsible for the alternation of double and single bonds between
two neighboring carbon atoms, a dimerized state \cite{lu}. As a
main characteristic of the SSH model Hamiltonian one notes that
the electron-phonon interaction modifies the electron hopping
matrix elements thus leading to a non-local (in momentum space)
coupling with vertex function dependent on both the electron and
the phonon wave vector. With the present paper we start an
investigation of the ground state and finite temperature
properties of systems which can be described by the SSH model.
Although some references to conducting polymers are made in the
introductive notation we focus here on the general features of
this  model which may well apply to a class of quasi one
dimensional systems with short range {\it e-ph} interactions
\cite{barisic,ovchi,schulz,acquarone}. As a preliminar goal we
evaluate the relevance of the mass renormalization (induced by
this non local type of electron-phonon coupling) which might
provide signatures of polaron formation in the system. In 1D, the
real space SSH Hamiltonian reads

\begin{eqnarray}
H=\,& & \sum_{r}J_{r,r+1} \bigl(f^{\dag}_r f_{r+1} +
f^{\dag}_{r+1} f_{r} \bigr) + \,
 \nonumber \\
& & \sum_{r}\Bigl({{p^2_r}\over {2M}} + {K \over 2}(u_r -
u_{r+1})^2
 \Bigr)\,
 \nonumber \\
J_{r,r+1}=\,& & - {1 \over 2}\bigl[ J + \alpha (u_r -
u_{r+1})\bigr] \label{1}
\end{eqnarray}

where $J$ is the nearest neighbors hopping integral for an
undistorted chain, $\alpha$ is the $\pi$ electron-phonon coupling,
$u_r$ is the dimerization coordinate which specifies the
displacement of the (CH) group on the $r-$ lattice site along the
molecular axis, $p_r$ is the momentum operator conjugate to $u_r$,
$M$ is the (CH) group mass, $K$ is the effective spring constant,
$f^{\dag}_r$ and $f_{r}$ create and destroy $\pi$-electrons on the
$r-$ (CH) group. Incidentally we note that, by means of a
Jordan-Wigner transformation, the Hamiltonian in eq.(1) maps onto
that of a spin-Peierls chain \cite{fradkin,zheng}. After expanding
the lattice displacement and its conjugate momentum in terms of
the phonon creation and annihilation operators $b^{\dag}_{q}$ and
$b_{q}$ and Fourier transforming the electron operators, one gets
the SSH Hamiltonian in momentum space:

\begin{eqnarray}
& &H=\, H_0 + H_{int}\,
 \nonumber \\
& &H_0=\, \sum_{k}\epsilon_k f^{\dag}_k f_k +  \sum_{q} \omega_q
b^{\dag}_q b_q \,
 \nonumber \\
& &H_{int}=\, \sum_{k,q} g(k+q,k) \bigl(b^{\dag}_{-q} + b_q \bigr)
f^{\dag}_{k+q} f_k \,
 \nonumber \\
& & \epsilon_k=\, -J \cos (k) \,
 \nonumber \\
& & \omega^2_q=\, 4{{K \over M}}\sin^2 ({q \over 2}) \,
 \nonumber \\
& &g(k+q,k)=\, {{i \alpha} \over {\sqrt {2MN \omega_q}}}
\bigl(\sin (k+q) - \sin (k) \bigr) \label{2}
\end{eqnarray}

where $N$ is the total number of lattice sites. The phonon
dispersion relation is defined in the range $q\in [0,\pi]$.
Assuming a reduced Brillouin zone ($|q| \le \pi/2$) the spectrum
displays both an acoustic and an optical branch. Here we attack
the Hamiltonian in eq.(2) by using the Matsubara Green's functions
formalism and taking the {\it e-ph} term as the perturbation.
Although a weak coupling perturbative approach cannot capture the
full multiphononic nature of the polaronic quasiparticle, an
envisaged, sizeable enhancement of the charge carrier mass  even
to the lowest order of perturbative theory would be an indicator
of polaron formation. The full electron propagator is defined as:

\begin{eqnarray}
G(p,\tau)=\,& &-\sum_{n=0}^\infty (-1)^n \int_0^\beta d\tau_1
...d\tau_n \Bigl < T_\tau f_p(\tau) H_{int}(\tau_1) \cdot \,
 \nonumber \\
 & &\cdot
H_{int}(\tau_n) f^\dag_p(0) \Bigr >_0 \label{3}
\end{eqnarray}

where $\beta$ is the inverse temperature, $T_\tau$ is the time
ordering operator, $<...>_0$ indicates that thermodynamic averages
are taken with respect to the unperturbed Hamiltonian and only
different connected diagrams contribute to any order $n$.

I have calculated the  self-energy terms due to one phonon ($n=2$
in eq.(3)) and two phonons ($n=4$ in eq.(3)) scattering processes.
Their finite temperature expressions are:

\begin{eqnarray}
\Sigma_p^{(1)}(i\varepsilon_m)=\,& &-\sum_q g^2(p,p-q) \Biggl[
{{n_B(\omega_q) + n_F(-\epsilon_{p-q})}\over {i\varepsilon_m -
\epsilon_{p-q} - \omega_q}} + \,
 \nonumber \\
 & &{{n_B(\omega_q) +
n_F(\epsilon_{p-q})}\over {i\varepsilon_m - \epsilon_{p-q} +
\omega_q}} \Biggr] \,
 \nonumber \\
\Sigma_p^{(2a)}(i\varepsilon_m)=\,& &{1 \over {\beta^2}}
\sum_{q,q_1} g(p,p-q)g(p,p-q_1) \cdot\,
 \nonumber \\
& & g(p-q,p-q-q_1)g(p-q_1,p-q-q_1) \cdot  \,
 \nonumber \\
& & \sum_{\omega_n,\omega_l}D^0_q(i\omega_n)D^0_{q_1}(i\omega_l)
G^0_{p-q}(i\varepsilon_m-i\omega_n)\cdot \,
 \nonumber \\
& &G^0_{p-q_1}(i\varepsilon_m-i\omega_l)
G^0_{p-q-q_1}(i\varepsilon_m-i\omega_n-i\omega_l) \,
 \nonumber \\
\Sigma_p^{(2b)}(i\varepsilon_m)=\,& &{1 \over {\beta^2}}
\sum_{q,q_1} g^2(p,p-q)g^2(p-q,p-q-q_1) \cdot  \,
 \nonumber \\
& & \sum_{\omega_n,\omega_l}D^0_q(i\omega_n)D^0_{q_1}(i\omega_l)
\Bigl[G^0_{p-q}(i\varepsilon_m-i\omega_n)\Bigr]^2 \cdot \,
 \nonumber \\
& &G^0_{p-q-q_1}(i\varepsilon_m-i\omega_n-i\omega_l) \,
 \nonumber \\
 \Sigma_p^{(2c)}(i\varepsilon_m)=\,& &{1 \over {\beta^2}} \sum_{q,k}
g^2(p,p-q)g^2(k,k+q) \cdot  \,
 \nonumber \\
& & \sum_{\omega_n}\Bigl[D^0_q(i\omega_n)\Bigr]^2
G^0_{p-q}(i\varepsilon_m-i\omega_n) \cdot \,
 \nonumber \\
& &\sum_{\varepsilon^{\prime}_m}
G^0_{q+k}(i\varepsilon^{\prime}_m+i\omega_n)G^0_k(i\varepsilon^{\prime}_m)
 \label{4}
\end{eqnarray}

There are three contributions due to different connected
two-phonons diagrams \cite{mahan}. $n_B(\omega_q)$ is the Bose
occupation factor  and $n_F(\epsilon_{p})$ is the Fermi occupation
factor. The frequencies $\varepsilon_m$ and $\omega_n$ are an odd
and an even multiple of $\pi/\beta$ respectively.
$G^0_{p}(i\varepsilon_m)$ is the free electron propagator and
$D^0_q(i\omega_n)$ is the free phonon propagator. By analytical
continuation $i\varepsilon_m \rightarrow \varepsilon + i\delta$
one gets the retarded self-energy whose real part determines the
renormalized electron mass $m_{eff}$:

\begin{equation}
{{m_{eff}}\over {m_0}}=\,{{1 - {{\partial
Re\Sigma_p(\varepsilon)}/ {\partial \varepsilon }}|_{p=0; \,\,
\varepsilon=-J}} \over {1 + {{\partial Re\Sigma_p(\varepsilon)}/
{\partial \epsilon_p }}}|_{p=0; \,\, \varepsilon=-J}} \label{5}
\end{equation}

where, $ Re\Sigma_p(\varepsilon)=\, Re\Sigma_p^{(1)}(\varepsilon)
+ Re\Sigma_p^{(2a)}(\varepsilon) +  Re\Sigma_p^{(2b)}(\varepsilon)
+ Re\Sigma_p^{(2c)}(\varepsilon)$, has been obtained from eqs.(4)
working out the straightforward but cumbersome double frequency
summations and taking the zero temperature limit. The model
contains three free parameters: the hopping integral $J$, the zone
boundary frequency $\omega_{\pi}=\,2\sqrt{{K/M}}$ which coincides
with the zone center  optical frequency in the reduced zone
scheme, the coupling constant $\alpha^2/4K$. In Fig.1, the mass
ratio of eq.(5) is plotted versus the adiabaticity parameter
$\omega_{\pi}/J$ assuming:(i) a narrow bare band $J$ value, (ii) a
weak {\it e-ph} coupling regime. $m_{eff}^{(1)}$ denotes the mass
renormalization due to the very one-phonon self-energy corrections
while $m_{eff}^{(2)}$ is enriched by the two-phonons scattering
processes. Particular care has to be taken in handling the
principal values which enter the real self-energy terms. I have
used the representation

\begin{equation}
P.P. \bigl({1 \over x}\bigr)=\,\lim_{\eta \to 0} {x \over {x^2 +
\eta^2}}$$
\end{equation}

and achieved numerical convergence by setting $\eta=\,10^{-4}$ and
summing over 200 $q$ points in each Brillouin zone. In the extreme
adiabatic regime, $Re\Sigma_p^{(1)}(\varepsilon)$ evaluated at the
band bottom is much larger than the two-phonons terms and the same
trend holds for the partial derivatives which enter the mass
ratio. Then, $m_{eff}^{(2)}$ does not differ essentially from
$m_{eff}^{(1)}$ and the renormalization is very poor since
scattering by low energy phonons does not enhance the electron
mass over the bare band value. As an example, at
$\omega_{\pi}=\,J/2$, we get (in units meV)
$Re\Sigma_0^{(1)}(-J)=\,1.45$, $Re\Sigma_0^{(2a)}(-J)=\,0.12$,
$Re\Sigma_0^{(2b)}(-J)=\,-0.17$, $Re\Sigma_0^{(2c)}(-J)=\,-0.5
\cdot 10^{-3}$. At $\omega_{\pi}\simeq \,3J/5$, the
$Re\Sigma_0^{(2a)}(-J)$ and  $Re\Sigma_0^{(2b)}(-J)$ terms are
comparable to  $Re\Sigma_0^{(1)}(-J)$ while the $2c$ diagram is
still negligible. Accordingly $m_{eff}^{(2)}$ starts to get larger
than $m_{eff}^{(1)}$ with an increase of $ \sim 15\%$ in the
intermediate $\omega_{\pi} \simeq\,J$ regime. Although, in this
regime, the application of a low order perturbative theory may be
questionable, the obtained mass enhancement  is likely a signature
that quasiparticles with polaronic character may form in the
system once multiphonons scattering processes become appreciable.
Incidentally we note that a variational study \cite{pucci} of the
SSH model finds in the intermediate regime favourable conditions
for the existence of localized polaronic solutions. Fig.2 displays
the one-phonon effective mass versus $\omega_{\pi}/J$ for two
values of {\it e-ph} coupling. Setting $J=\,50meV$ we take here an
electron band narrower than in Fig.1. Again, the mass increase is
relevant only in the intermediate $\omega_{\pi}/J$- range  while
the electrons are essentially free in the extreme adiabatic and
antiadiabatic regimes. The spike at $\omega_{\pi} \sim \sqrt{2} J$
which dominates the mass behavior is mainly due to scattering by
$|q|=\,\pi/2$-phonons. This feature is related to the 1D electron
and phonon dispersion relations and it may be partly suppressed in
higher dimensionality. In the $\omega_{\pi}/J$- window which is
sensitive to renormalization effects  the {\it e-ph} coupling
parameter slightly affects the values of the one-phonon effective
mass. For comparison also the one phonon  effective mass of the
Holstein-like model is reported on in Fig.2. To point out the role
of the {\it e-ph} coupling we have replaced the $g^2(p,p-q)$
function of the SSH model by a constant $g_H^2$ without any change
to the dispersion relations. Then the Holstein-like model here
assumed differs from the "true" Holstein model \cite{holst59} of
diatomic molecules in which purely optical phonons couple to the
electrons. We also note that simplified models with dispersionless
Einstein phonon spectra would lead to wrong estimates of the
ground state properties \cite{io1,capone}. Setting
$g_H^2=\,\alpha^2/4K \cdot \omega_{\pi}$ with $\alpha^2/4K=1meV$
we obtain a mass behavior similar to that of the SSH model: again
we find an abrupt mass increase at $\omega_{\pi} \sim \sqrt{2} J$
(thus confirming its dependence on the 1D dispersion relations
rather than on the choice of the coupling function) but the
present mass values are generally lower than those predicted by
the Holstein model in strong coupling perturbative theory
\cite{io3}. This suggests that the well known mass enhancement of
the Holstein adiabatic and antiadiabatic polaron is due to the
multiphononic effects fully captured (for instance) by the
Lang-Firsov \cite{lang} strong coupling method whereas the details
of the short range coupling seem to have little influence on zone
center properties such as the effective mass. Conversely one might
evaluate the SSH polaron mass in a strong coupling approach to
check whether and to which extent the momentum dependent vertex
function plays there a peculiar role. Finally, Fig.3 emphasizes
that the effective coupling $\alpha^2/4K$ scarcely affect the mass
renormalization in adiabatic and intermediate regimes while a
slight mass dependence on $\alpha^2/4K$ is observed in the
antiadiabatic regime with very narrow electron band
$J=\,\omega_{\pi}/2$. In the latter case however, in the upper
portion of the $x$ axis, the dimensionless coupling
$\alpha^2/(4KJ)$ is larger than one and the perturbative method
breaks down. In the intermediate case $\omega_{\pi}=\,J$ also the
two-phonons mass is reported on.

In conclusion, I have assumed a weak coupling perturbative regime
and applied low order diagrammatic techniques to the one
dimensional Su-Schrieffer-Heeger Hamiltonian in order to compute
the electron mass renormalization induced by one- and two-phonons
scattering. Tuning the parameter $\omega_{\pi}/J$, the mass
behavior has been analysed both in adiabatic, intermediate and
antiadiabatic conditions. There is a sizable mass enhancement only
in the intermediate and moderately antiadiabatic range whereas the
electrons don't drag phonons whose energy is either much smaller
or much larger than the electron energy. Hence, no mass
enhancement is found in the extreme adiabatic and antiadiabatic
regimes. Replacing the momentum dependent vertex function by an
Holstein-like coupling constant does not modify substantially the
effective mass due to the one phonon self-energy diagram. We
emphasize that this result holds in the present weak coupling
perturbative approach while different conclusions may be drawn in
strong coupling theories. The two phonons self-energy corrections
introduce an appreciable mass enhancement mostly in the
intermediate $\omega_{\pi} \sim J$ regime. This  feature can be
likely interpreted as an onset of polaronic crossover whose entity
and precise location in parameter space requires however
computations of multiphonons effects.

\begin{figure}
\vspace*{8truecm} \caption{Renormalized masses (in units of bare
band electron mass) versus the adiabaticity parameter.
$m_{eff}^{(1)}$ is due to the one phonon self-energy correction.
$m_{eff}^{(2)}$ includes also the two phonons self-energy terms.}
\end{figure}

\begin{figure}
\vspace*{8truecm} \caption{Renormalized masses (in units of bare
band electron mass) as due to the one phonon self-energy
correction versus the adiabaticity parameter. Two values of
electron-phonon coupling have been chosen. The mass behavior
obtained in a Holstein-like model with constant coupling is
reported on for comparison. }
\end{figure}

\begin{figure}
\vspace*{8truecm} \caption{Mass renormalization (in units of bare
band electron mass) due to one phonon scattering  versus the
electron-phonon coupling (in meV). Three values of the
adiabaticity parameter have been chosen. In the intermediate case
$J=\,\omega_{\pi}$ also the effect of the two phonons self-energy
correction is displayed. }
\end{figure}

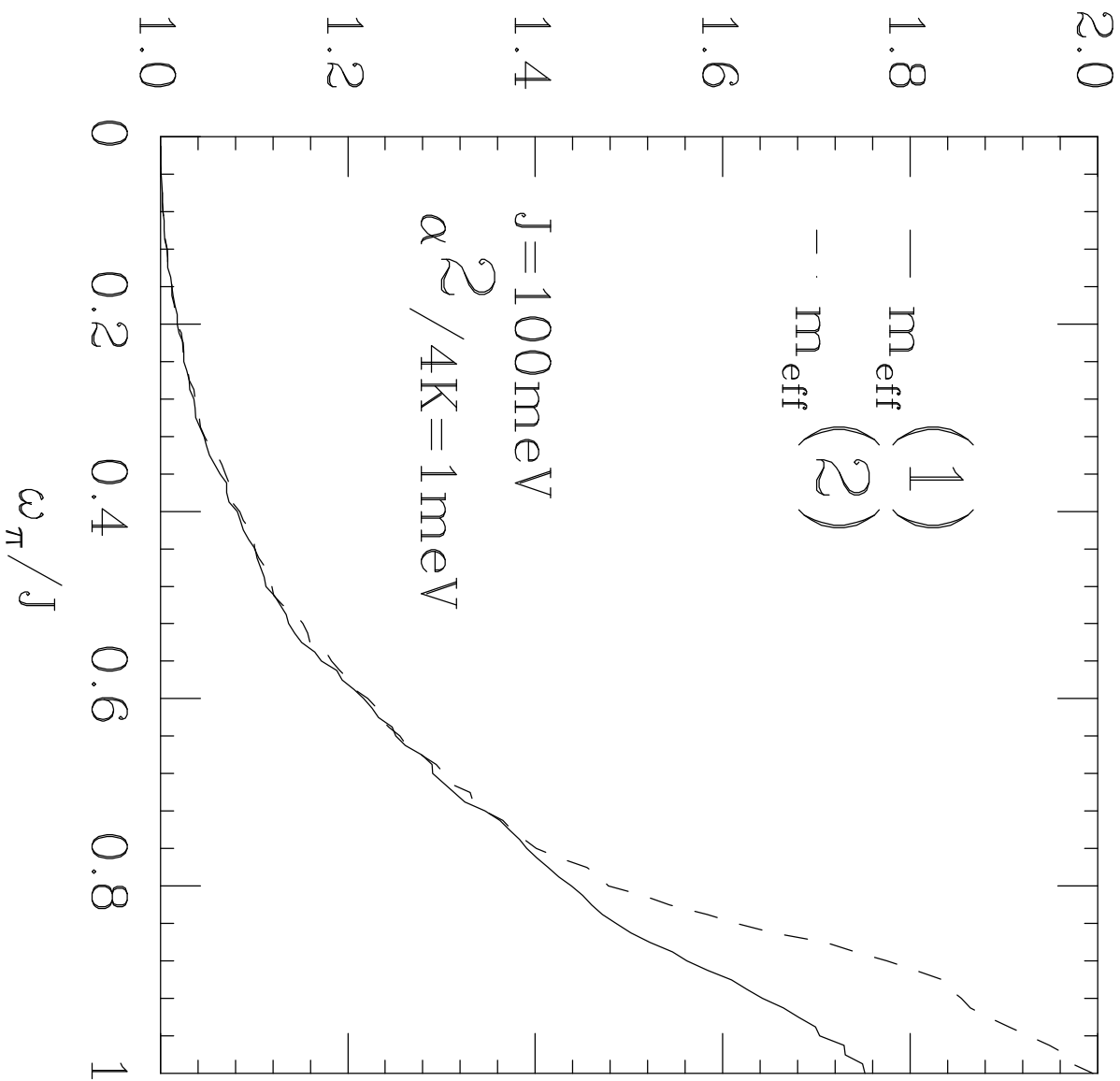
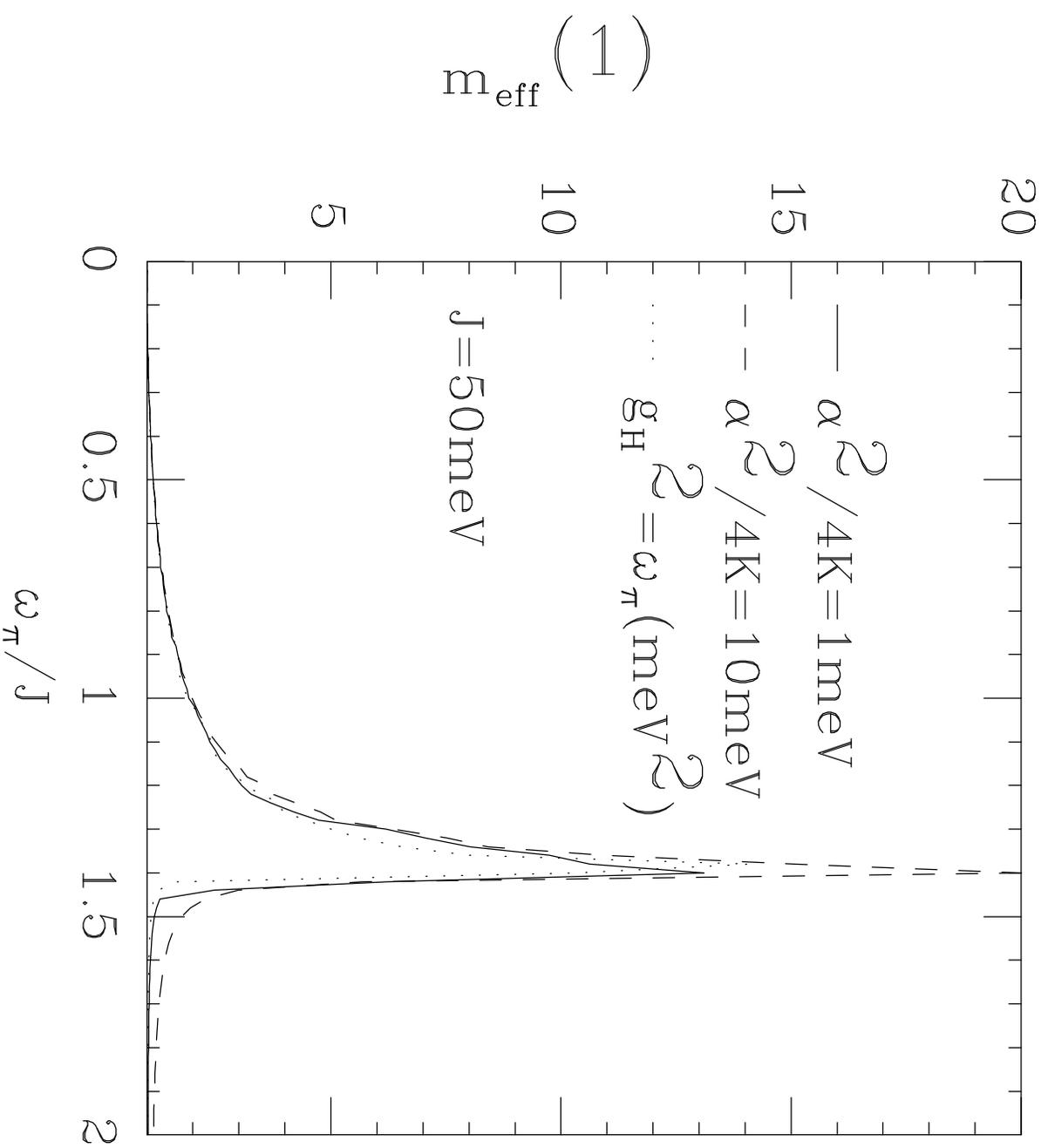
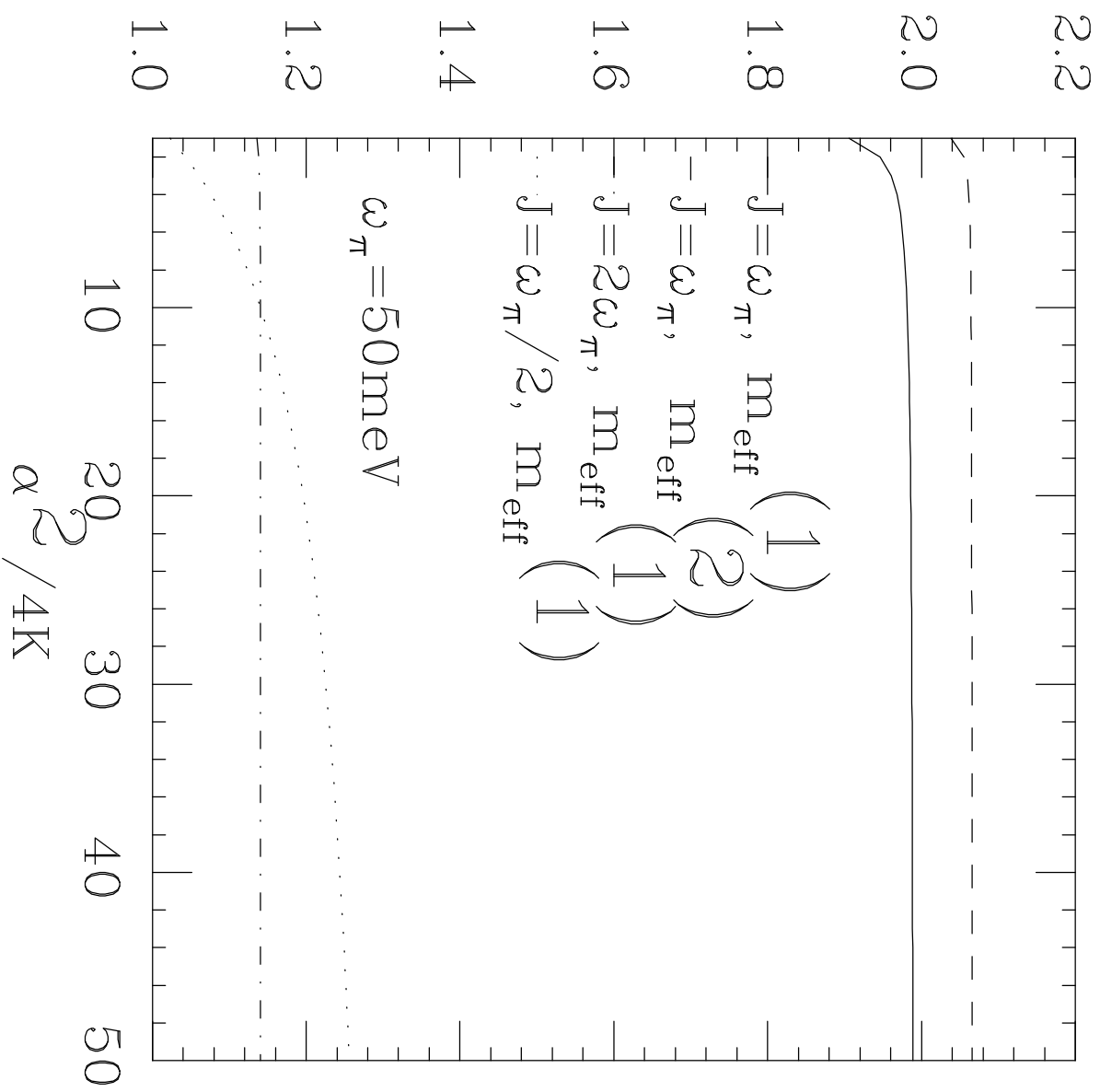
\end{document}